 \documentstyle[12pt,draft,epsf]{article}
 \newcommand \be {\begin{equation}}
\newcommand \bea {\begin{eqnarray} \nonumber }
\newcommand \ee {\end{equation}}
\newcommand \eea {\end{eqnarray}}
 
 \newcommand \s {\sigma}

\newcommand \g {\gamma}
\newcommand \la {\lambda}

 \newcommand \al {\alpha}

\newcommand \LL{{\cal L}}
\newcommand \Ht{{\cal H}_t}

\newcommand \lan {\langle}
\newcommand \ran {\rangle}
\newcommand \pa {\partial}
\newcommand \bi {\bibitem}
\newcommand \bit {\begin{itemize}}
\newcommand \eit {\end{itemize}}
\newcommand \sign {\mbox{sign}}

\begin{document}

\title{On the Statistical Properties of the Large Time Zero Temperature
Dynamics of the SK Model}
 \author{  Giorgio Parisi \\
Dipartimento di Fisica, Universit\`a {\sl La  Sapienza}\\
INFN Sezione di Roma I \\
Piazzale Aldo Moro, Roma 00187}
\maketitle

\begin{abstract}
In this note we study the zero temperature dynamics of the Sherrington
Kirkppatric model and we investigate the statistical properties of the
configurations that are obtained in the large time limit. We find that the
replica symmetry is broken (in a weak sense). We also present some general
considerations on the synchronic approach to the off-equilibrium dynamics,
which have motivated the present study.
 \end{abstract}
\vfill
\vfill
\newpage

\section {Introduction}
In recent years there have been many progresses in our understanding of the
non-equili\-brium dynamics   \cite{BO1}-\cite{FERRARO} of the infinite
Sherrington Kirkpatrick model \cite{MPV,PB2} and  of other  glassy systems
\cite{FM,CUKU2,MPR}. The aim of these studies is to compute the  properties of
the systems (e.g. the energy or the magnetization) as function of time if we
know the  initial configuration at
time zero.

These progresses have been done using a diachronic approach in which the
evolution of the
system is studied by writing closed equations for the correlation functions and
response functions at
different times.

An alternative synchronic approach has been put forward \cite{CS,CF}; here we
consider the
probability distribution of the configurations of the system at a given time
$t$ ($P_t(\s))$ and we
write it as
\be
P_t(\s) \propto \exp(- \Ht(\s)).
\ee

In this approach there are two crucial steps:
\begin {itemize}
\item{(a)}
The determination of the effective Hamiltonian $\Ht$. It is possible that in
order to obtain
qualitative and semiquantitative information it is not necessary to compute
exactly $\Ht$, and an
approximate knowledge is sufficient.
 \item{(b)}
The computation of the statistical properties of the system at given effective
Hamiltonian.
\end{itemize}

This approach may be successfully only if the effective Hamiltonian $\Ht$ (or a
reasonable approximation to it) is not too complicated. While  the diachronic
approach is  rather systematic, a good amount of guesswork is needed  in the
synchronic approach in order to  chose a reasonable form of the effective
Hamiltonian.

In order to get some intuition on the possible forms of the effective
Hamiltonian we have studied in this note the statistical properties of the
configurations that are obtained at  large times in the SK model using a zero
temperature dynamics. This problem has already been  studied in the past
\cite{OPPER,FERRARO,PPV,PP,PARGA,MP}, and it has its own interest. We have
obtained  some new and unexpected results, i.e. we have found that the replica
symmetry is broken (in  a weak sense \cite{PV,P1}) and that the distribution
of the local fields has uneexpected properties.

In the second section we present some general considerations on the synchronic
approach. These considerations have been the motivation of the present study,
but they  may be skipped by the reader interested only in the results of the
paper for the SK model. In the third section we define the dynamics which we
study and we recall some  known results on the zero temperature solutions of
the TAP equations \cite{TAP}. In section IV we  present the numerical results
of our investigations and finally in the last  section we present some
tentative conclusions.

\section {The Synchronous Approach}

We consider a system whose variables satisfy some kind of stochastic or
deterministic evolution equations.The probability distribution of the
configurations at time  zero is given:
\be
P_0\propto \exp (-{\cal H}_0).
\ee
The quantity ${\cal H}_0$ plays the role of boundary condition.

In the simplest case  the system is random at time zero and  ${\cal H}_0 =0$.
 Our aim is to compute  $\Ht$ given ${\cal H}_0$. If we are rather  luckily we
can compute it exactly. This happens for  example in the Gaussian SK \cite{CF}
or in the spherical SK model. Before discussing the general  approach it may be
useful to show in details the soluble example.

\subsection{The Gaussian Sherrington Kirkpatrick model}

In the  Gaussian Sherrington Kirkpatrick \cite{CF} the Hamiltonian is
\be
H= - \sum_{i,k} J_{i,k} \s_i\ \s_k, \label{SK}
\ee
where the variables $\s_i$ are real, with an intrinsic probability
distribution  at infinite temperature equal to the product of uncorrelated
Gaussian of unit variance.

 The  variables $J_{i,k}$ are usually Gaussian
distributed and uncorrelated. The precise form of the  distribution of
variable $J$ is not important here; we will only assume for simplicity of
notations that the  matrix $J$ does not have zero eigenvalues.

The dynamic is given by the Langevin equation:
\be
{d\s_i \over dt} = - \s_i +\beta F_i(t) +\eta_i(t), \label{LAN}
\ee
where the force is given by
\be
 F_i(t)=- {\pa H \over \pa \s_i}= \sum_k J_{i,k} \s(t)_k.
\ee
 The variables $\eta$ are have an uncorrelated white noise distribution
\be
\lan \eta_i(t) \eta_k(t')\ran = 2 T \delta(t-t') \delta_{i,k}.
\ee

We suppose that at time zero the initial condition is simply $\s(0)_i=0$.
It is easy to show that the effective Hamiltonian at time $t$ is given by
\be
\Ht(\s) = \sum_{i,k} \s_i f^t(J)_{i,k} \s_k,
\ee
where the function $f^t$ is given by
\be
f^t(z) ={z \over 1- \exp(-tz)} \label{MAGIC}
\ee
and $f^t(J)_{i,k}$ denote the $i,k$ element of the matrix $f^t(J)$.

The proof of this statement may be obtained by noticing that the evolution
equations becomes much simpler in the basis where the matrix $J$ is diagonal
\cite{CF,CKP,CU}. In  this basis it is easy to see that the components of the
variables $\s$ along the eigenvectors of  the matrix $J$ are uncorrelated.
Their variance can be computed and in this way one obtains
equation (\ref{MAGIC}).

\subsection{The general approach}

In the general case the system has an Hamiltonian $H_J(\s)$ and the evolution
is described by a Langevin equation of the form in eq. (\ref{LAN}) or by some
analogous equation for  discrete systems.

We suppose that the effective Hamiltonian can be written as
\be
{\cal H}_0 = g(\s,J,\la(t)),
\ee
where $g$ is a preassigned function which depend on  $M$ variables
$\la_\al(t)$ ($\al=1...M$); $M$ may also be  infinite. In this framework
the effective Hamiltonian depends on time only trough the variables $\la(t)$.

If we suppose to know the function $g$ (or a good approximation to it), the
problem consists in finding the appropriate values of the functions
$\la_\al(t)$.

There  are two  strategies which we can follow:
 \begin{itemize}
\item (a)
 We choose a set of observables $O_\al(\s,J)$ for $\al=1...M$ and we impose
the validity of the equations:
\be
  {d\lan O_\al\ran_t \over dt}=
 \sum_{\g} {\pa \lan  O_\al \ran_t \over \pa \la_\g}{d \la_\g \over dt},
\ee
where the r.h.s. is computed using the Langevin equatios.

If we are new to equilibrium it is convenient to write
\be
\Ht = \beta H + \sum_{\g}  O_\g \la_\g(t).
\ee
For small values of $\la_\g(t)$ we can linearize the equations and we get
\be
\sum_\g \lan O_\al O_\g \ran_c {d \la_\g(t) \over dt}=
\sum_\g  \lan \sum_i {\partial O_\al \over \partial s_i}
 {\partial O_\g \over \partial s_i} \ran \la_\g(t),
\ee
where the expectation values are computed at equilibrium and $\lan \ran_c$
denotes the connected expectation value.
\item (b)
We write down the Fokker Plank equation:
\be
{dP \over dt}= \LL P(T),
\ee
where $\LL$ is  the appropriate linear operator.
A variational principle may be used  to chose the variables $\la_n(t)$. For
example  we can impose that
\be
\sum_{\{\s\}}(\sum_{\g} {\pa P(t) \over \pa \la_\g}{d \la_\g \over dt} -\LL
P(t))^2
\ee
takes the minimum value.
\end{itemize}

In the past this point of view has been advocated in the study of turbulence
and both strategies have  been followed
\cite{BBP,MIG}.

In their original work Cooley and Sherrington \cite{CS} have analysed the
dynamics of
the SK model with Ising spins. They have taken $M=2$. In the case of all
spins equal to 1 at time
zero, they have
 \be
g(\s,J,\la(t))= \la_1(t) \sum_{i,k}J_{i,k}\s_i\s_k +\la_2(t)\sum_i \s_i
\ee
Reasonable results have been obtained expecially at the short times.

It is clear that the previous equation (with (M=2)) is only a first
approximation. In the case of the Gaussian SK model it was   proved \cite{CF}
that it does not reproduce the exact  result. Indeed in this last case we have
seen that the correct expression (for $\s_i(0)=0$) is  given by
 \be
g(\s,J,\la(t))= \sum_{\al=1,\infty} \la_\al(t)
\sum_{i,k}(J^{\al})_{i,k}\s_i\s_k,
\ee
where as usual $(J^\al)_{i,k}$ is the matrix element of the matrix $J$ to the
power $\al$.

For the Ising case is possible that the previous formula is not adequate and
that other terms may be
present as
\be
\sum_i \s_i (\sum_k J_{i,k}\s_k)^3.
 \ee

It is quite reasonable that near to the critical temperature too high powers of
the variables
$\s$ are not present so that it possible that a manageable approximation may be
done using the
techniques developed in \cite{MPR}.

\section{Zero temperature dynamics and the TAP equations}

Here we consider the SK Ising model with the Hamiltonian in eq. (\ref{SK}),
with  the $J$ random independent Gaussian variables with variance ${1\over
N}$, $N$ being the total  number of Ising spins. We study later the case where
the variables $J$ take the values $\pm  N^{-1/2}$. The two models coincide in
the limit $N \to \infty$.

We are interested in studying the statistical properties of the configurations
that are obtained by the dynamics at zero temperature at large times. In this
case the dynamics is such to orient  the spins with the effective field
$h_i=\sum_k J_{i,k} \s_k$. In other words one set
  \be
\s_i(t+1)= \sign(h_i(t)).\label{RULE}
\ee
Different algorithms differs in the order in which the rule eq. (\ref{RULE}) is
applied.

The algorithm stops when in all the sites the following equation is satisfied
\be
\s_i(t)= \sign(h_i(t)).\label{TAPEQ}
\ee

The asymptotic configuration at large times depend on the initial
configuration. We are interested in
studied the ensemble in which the each solution of zero temperature TAP
equation (\ref{TAPEQ})
is weighted with the probability of being obtained by the algorithm with a
random choice of the
initial configuration. In other words we weight each solution with the size of
its attraction basin.

It is known that in the large $N$ limit the energy density of the asymptotic
configuration does not
depend on the stating configuration with probability one and it is higher that
the ground state
energy. Indeed for a sequential algorithm it is about $E_S=-.715$ \cite{PP},
while the ground state
energy is $E_0=-.7633$.

The simplest hypothesis would be that the set of configuration weighted with
the attraction basin statistically coincide with the set of all solutions of
the TAP equation with  energy equal to $E$.

A precise computation of the
statistical properties of the solutions of the zero temperature TAP
equations for this value of the energy has not been done.
However  we  can  use the information we have  for energies greater that
$E_{RSB}=-.672$ (at which an  exact computation can be done) and  for energy
equal to the ground state the value. The energy $E_S$ is  intermediate among
the two so that an educated guess can be done. We could guess that the
maximum  overlap  among two generic solutions  \footnote{We will define this
quantity with greater precision later on.}  is is 0 at $E_{RSB}$ and it is 1
at $E_0$ so that we could guess its value around .6-.7.
In general  the total number of solutions increases about as $\exp(AN)$, where
$A=0$ at $E_0$ and $A=.12$ at $E_{RSB}$. We guess a value of $A$ in the range
$.04-0.06$. These two values are  purely indicative.

The probability distribution of the effective field $P(h)$ is a shifted
Gaussian for energies
greater that $E_{RSB}$. There are no indications that $P(0)$ should be zero for
$E <E_{RSB}$.

We will see later that some these expectations are in variance with our
results coming from numerical simulations. We conclude that the original
hypothesis is wrong and that the  generic solution of the TAP equations,
weighted with its attraction basin, is not the generic solution  of the TAP
equations of the appropriate energy.

\begin{figure}
  \epsfxsize=400pt
\epsffile[22 206 649 650]{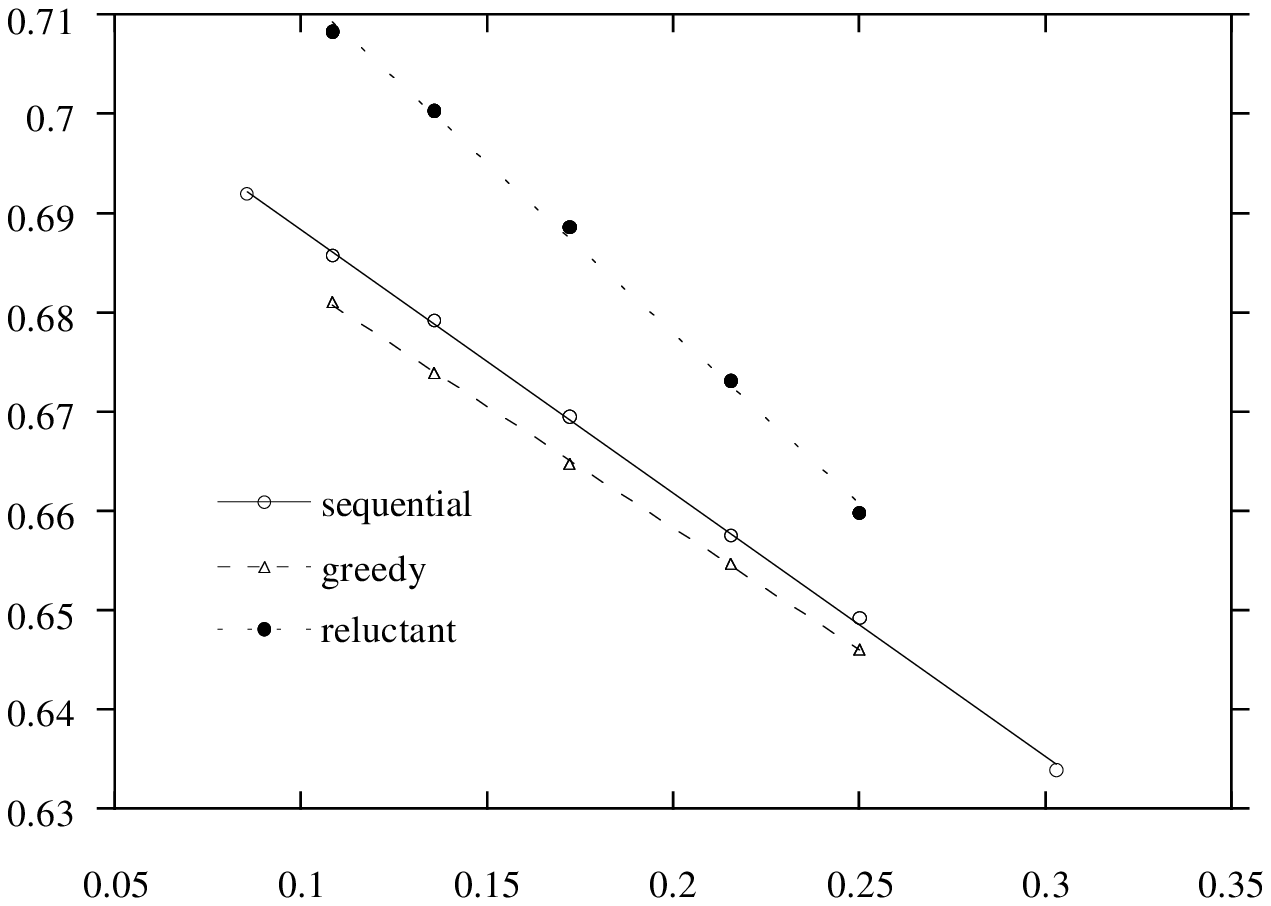}
  \caption[a]{\protect\label{UNO} Minus the average energy as function of
$N^{-1/3}$ for the sequential, greedy and reluctant algoritms. The lines are
linear fits.
  }
\end{figure}

Before presenting the numerical results we will describe the three minimisation
algorithms that we
have used: the sequential algorithm, the greedy algorithm and the reluctant
algorithm.
\bit
\item The Sequential Algorithm.

This is the simplest algorithm to implement. One cycle of the algorithm
consists in applying the rule
(\ref{RULE}), sequentially for increasing $i$ from $i=1$ to $i=N$. We repeat
the cycles up to the
moment at which a solution of the TAP equation is reached.
 This algorithm corresponds to the zero temperature limit of an usual Monte
Carlo or heath bath
dynamics.

\item The Greedy Algorithm.

This is the simplest algorithm to understand analytically. One step of the
algorithm consists in applying the rule (\ref{RULE}), for that $i$ which
minimises $\s_i h_i$. We  repeat the steps up to the moment at which a
solution of the TAP equation is reached.
 This algorithm corresponds to the zero temperature limit of the Glauber
dynamics.

\item The Reluctant Algorithm.

This  algorithm is the opposite of the greedy algorithm. One step of the
algorithm consists in applying the rule (\ref{RULE}), for that $i$ which
maximises $\s_i h_i$ among  those $i$ such that  $\s_i h_i$ is negative. One
repeats the steps up to the moment at which a solution of the TAP equation is
reached.
 At each step the energy decreases, but it decreases of the smallest possible
amount. As we shall see later this algorithm is the most effective in finding
the  configurations of smallest energies.
 \eit
\begin{figure}
  \epsfxsize=400pt
\epsffile[22 206 649 650]{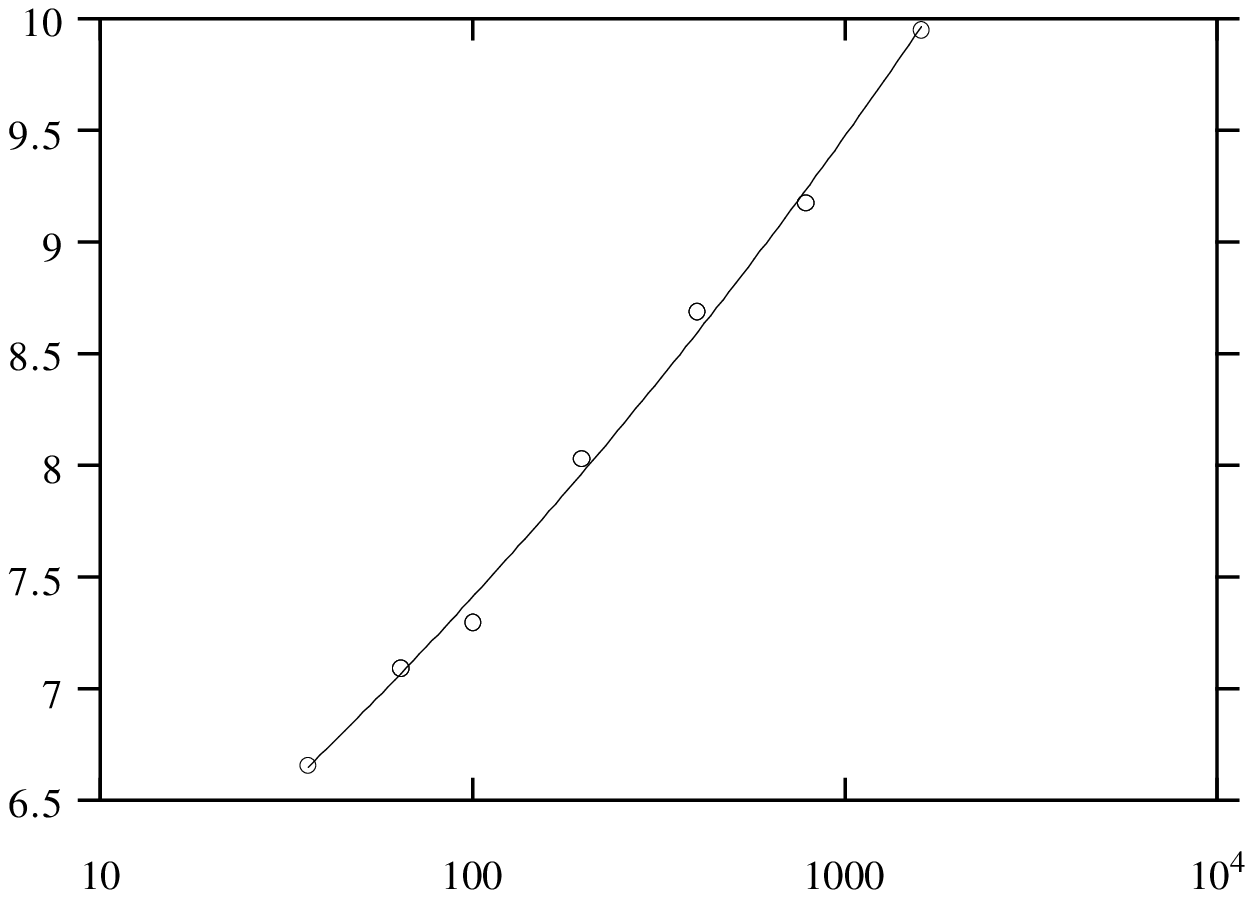}
  \caption[a]{\protect\label{DUE} The  average value of $ N \lan q^2\ran$ as
function of $N$ in the case of the sequential algorithm. The line is a power
 law fit with exponent .1; a logarithmic dependance (a straight line) is also
possible.
  }
\end{figure}

\section{Numerical Results}

  We have studied numerically the zero temperature dynamics of the SK model
for  systems with $N$ in the range 16-1600. We report firstly the results for
sequential updating; the results for the other
algorithms (the greedy and the reluctant) are not  qualitatively different
and they will not be discussed in details.

For
each value of $N$ many instances of the system have been generated (for 10000
at $N=16$ to $100$ at
$N=1600$). For each choice of the coupling $J$ we have followed the dynamics
starting from
$N^{1/2}$ different  intitial configurations. For each configuration we have
recorded the energy,
the distribution of the forces $h_i$; we have also compute the overlap
\be
q= |{\sum_{i=1,N} \s_i \tau_i \over N}|
\ee
among all the $N^{1/2}$ final configurations.

In fig. (\ref{UNO}) we show the expectation value of the energy density $E$ as
function of  $N$. The data are has been fitted as
\be
E= E_\infty + {C \over N^\al},
\ee
where the values of the parameters of the fit are $ E_\infty =-.715$, $C=.25$
and  $\al =.33$. The
value of  $ E_\infty $ differs from the gound state energy (which is -.7633).
The value
of $\al$ is similar to the one which is obtained for the $N$ dependence of the
ground state and it is compatible with being equal to $1/3$.

 The fluctuations of the energy density  go to zero when
$N$ goes to infinity  approximately
proportionally to $1/N$ (a best fit gives $1/N^{.98}$).

The expectation value of $q$ as function of $N$ goes to zero approximately as
$N^{-.9}$ when $N
\to \infty$ (see fig. (\ref{DUE}).).  A behaviour of the type ${\ln(N) \over
N}$ cannot be excluded.

\begin{figure}
  \epsfxsize=400pt\epsffile[22 206 649 650]{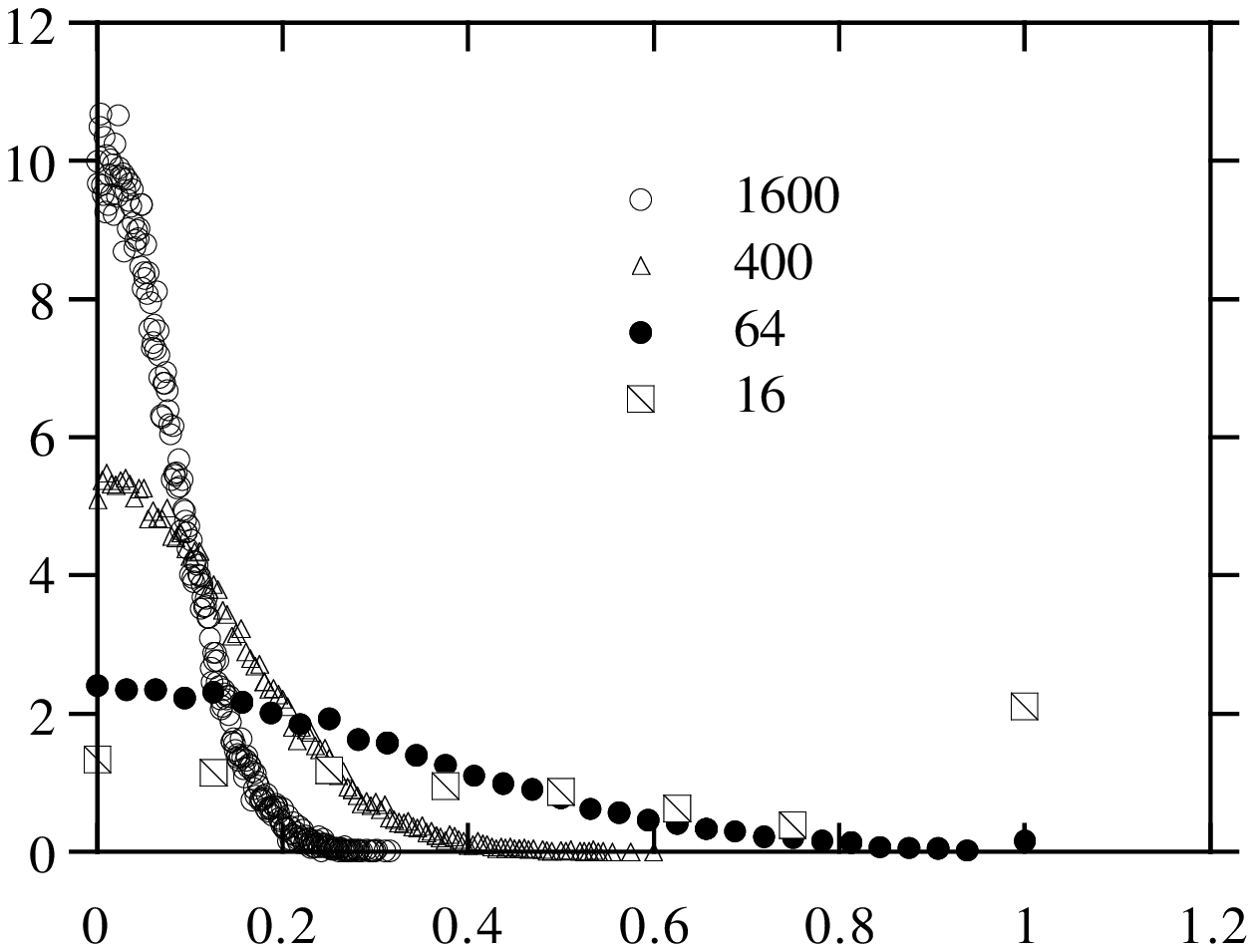}
  \caption[a]{\protect\label{SEI} The function $P(q)$ for different values of
$N$ ($N=16,64,400,1600$) in the case of the sequential algorithm.
  }
\end{figure}

These results are in good agreement with older studies \cite {PP,PARGA}. The
surprise come from the study of the distribution function $P(q)$. If we plot
the function $P(q)$  for diffeent sizes we do not see anything unexpected (see
fig. (\ref{SEI})). The distribution becomes more and  more concentrated at
$q=0$ when $N \to \infty$.

The crucial point is however how fast the function $P(q)$ decreases with $N$
at fixed $q$.  More precisely  can define
the function
\be
f(q)= - \lim_{N \to \infty} {\ln(P_N(q)) \over N}.
\ee

\begin{figure}
  \epsfxsize=400pt\epsffile[22 206 649 650]{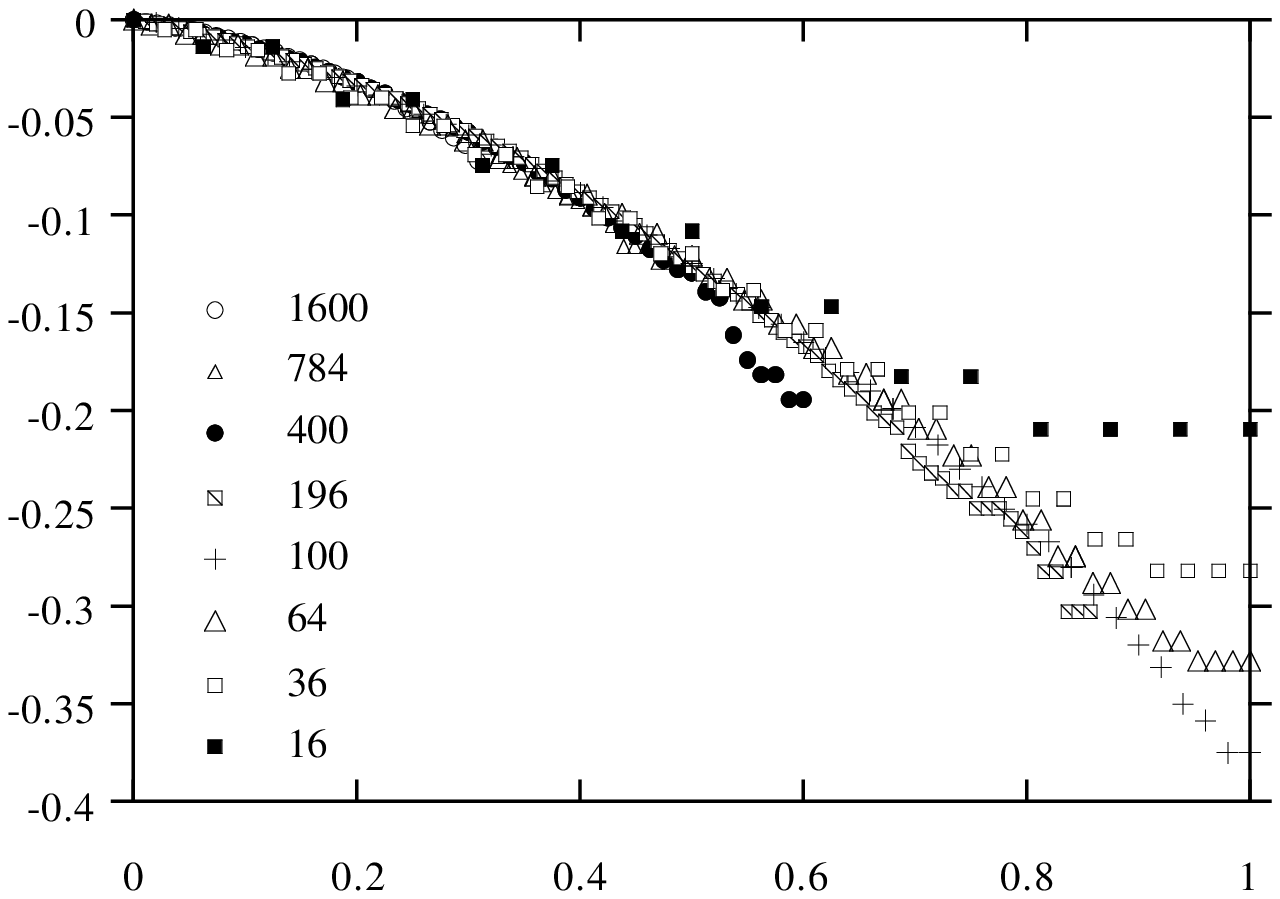}
  \caption[a]{\protect\label{QUATTRO}
    The results obtained with the sequential
algorithm for the function $r_N(q)$
for $N$ and $q$  ranging from 16 to 1600 and from 0 to 1 respectively.
  }
\end{figure}

In the ususal equilibrium SK model the function $f(q)$ is equal to zero in the
interval [0-$q_{EA}$]
and it is  different from zero for $q>q_{EA}$.

The region where $f(q)$ is zero  is the essential support of the function
$P(q)$. In the region where $f(q)=0$ we can modifying the  Hamiltonian by a
quantity  whose density goes to zero with $N$ (e.g. imposing the appropriate
bondary  conditions) and we can obtain a system such that the value is $q$.

If the function $P_N(q)$ becomes non trivial in the limit $N \to \infty$ the
replica symmetry is broken in the usual sence. On the contrary, if the
function  $P_N(q)$ becomes a delta function in the limit $N$ going to infinity,
but $f(q)$ is zero in a finite  interval, the replica symmetry is broken in a
weak sence.
\begin{figure}
  \epsfxsize=400pt\epsffile[22 206 649 650]{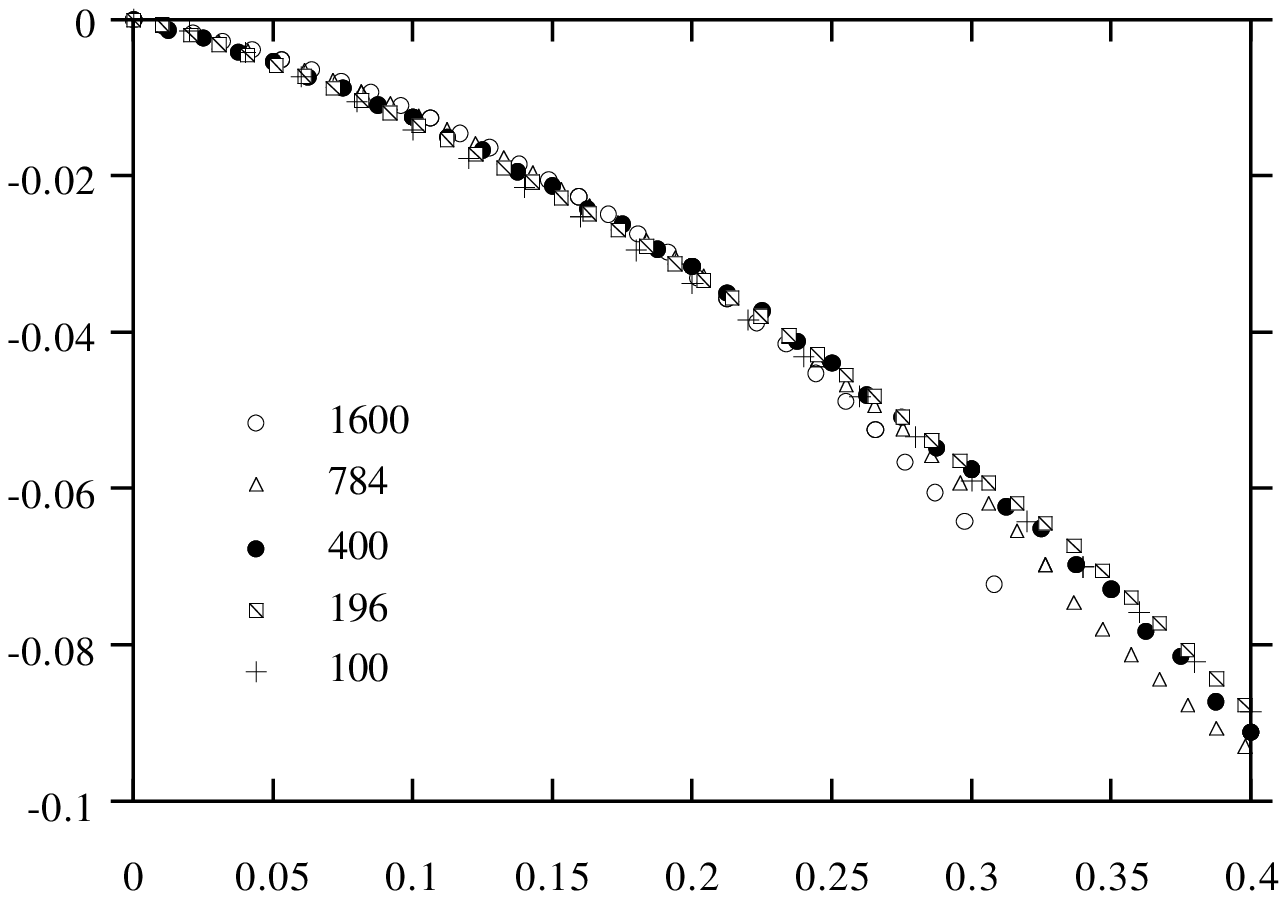}
  \caption[a]{\protect\label{CINQUE}
     The results obtained with the sequential algorithm for the
 function $r_N(q)$ for $N$ and $q$  ranging from 100 to 1600 and from 0
to .4 respectively.
  }
\end{figure}

 In fig. (\ref{QUATTRO},\ref{CINQUE}) we plot the quantity
\be
r_N(q)= N^{-\nu}\ln(\int_q^1 dq'P_N(q')) ,
\ee
with $\nu=2/3$.

This quantity has a very weak dependence on $N$ and it is rather likely that it
goes to a non zero value when $N$ goes to infinity in the whole interval 0-1.
If we accept this conclusion, we find that  $P_N(q)$ decreases slower than an
exponential for all values of $q$, included $q=1$ and therefore  the function
$f(q)$ is always zero in the whole interval 0-1. Replica symmetry is
thus broken  in a weak sence.

If $r_N(1)$ has a finite limit, as it is suggested by this data, the
probability of finding two
times the same solution goes to zero slower than an exponential of $N$,
implying that the effective
numer of different solutions which can be reached with this method diverges
slower that
an exponential of $N$.

The other quantity for which we obtain surprising results is the probabily
distribution of the fields $h_i=\sum_k J_{i,k} \s_k$. The function $P_N(h)$
depends weakly on $N$  (the expectation value of $h$ is the energy.

\begin{figure}
  \epsfxsize=400pt\epsffile[22 206 649 650]{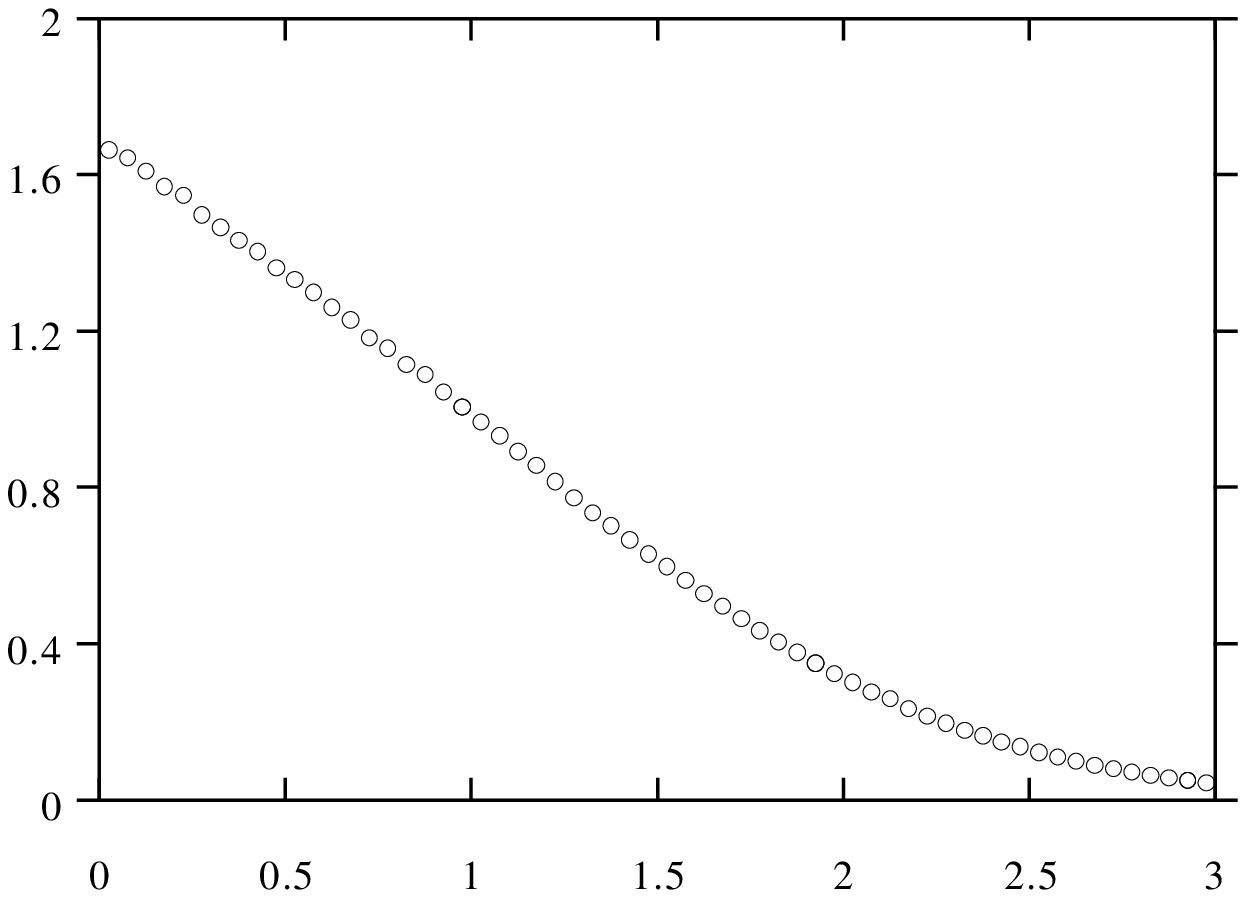}
  \caption[a]{\protect\label{TRE}
    The function $\tilde P(h))$  in the case of the sequential algorithm.}
\end{figure}

 The
function $P_N(h)$ seem to vanish linearly at $h=0$. In order to show this
behaviour in fig. (\ref{TRE}) we plot the function
\be
\tilde P_N(h) \equiv {P_N(h) \over  h+C},
\ee
for $N=1600$ where the constant $C$ (which vanishes when $N \to \infty$) has
been choosen in such a way to have a smooth behaviour at small $h$. We have
found that $C= 2 N^{-1/2}$ is a good choice. This value of $C$ is natural,
indeed the variable $h$ make take values $(n+1/2)C$, with $n$ integer.

A linear behaviour of the function $P(h)$ is similar to the one observed at
thermal equilibrium at zero temperature, where $P(h)$ is proportional to $h$.

We do not have a simple explanation for such behaviour. The process of
minimizing the energy of a given spin has the side effect of decreasing also
the energy of the  other spins, pushing the distribution of $h$ far from zero.
Howevever it is unclear how to  transfom this observation in a quantitive
prediction.

We have analized also the other dynamics ad we have obtained comparable
results. The only remarkable effect that  the dependence of the energy on the
algorithm is counterintuitive. Indeed the greedy algorithm is  the worst (the
energy is higher of about .007 than the sequential algorithm) and the reluctant
algorithm is much better (the energy -.746 is lower of about .03 than the
sequential algorithm).  This effect is likely due to the fact that the  greedy
algorithm is more easily trapped in local minima with high energies and that
the slower reluctant  algorithm avoids these traps.

  \section{Conclusions}

The probability distribution of the configurations obtained by the zero
temperature dynamics have rather peculiar statistical properties. The results
obtained are very different  from those coming taking a generic solution of
the TAP equations with the appropriate energy. The  maximum allowed value of
$q$ (from the thermodynamic point) is 1, not smaller than $1$ as it  would
happen for the generic solutions of that energy. Morover the distribution of
$h$ is zero at  $h$ =0, while there are no reasons that $P(h)$ should be equal
to zero at $h=0$ in the case if the  generic solution.

 It is rather likely the  corresponding effective Hamiltonian is not so simple
and its  statistical properties should be computed using a replica symmetry
breaking approach.

It would also interesting to connect the small $h$ behaviour with the power
laws observed in the dependence of physical quantities (like the energy of the
remanent  magnetization) or the number of iterations in the sequential
dynamics.

 \section*{Acknowledgements}
It is a pleasure to thank E. Marinari for useful discussions.


\begin{thebibliography}{99}

\bi{BO1} J.-P. Bouchaud; J. Phys. France {\bf 2} 1705, (1992).

\bi{PR} G. Parisi and F. Ritort , J. Phys. {\bf A 26} 6711 (1993).

\bi{CUKURI} L. F. Cugliandolo, J. Kurchan and F. Ritort, Phys. Rev. {\bf
B49},  6331 (1994).

\bi{OPPER} Phys. Rev. Letters {\bf 71}, 71, (1992).

\bi{CUKU1} L. F. Cugliandolo and J. Kurchan; J. Phys. {\bf A27} (1994) 5749;
cond-mat {\bf 9403040}, Phil. Mag. (to be published).

\bi{FERRARO}G. Ferraro, Rome Preprint (1994)

 \bibitem{MPV}  M.~M\'ezard, G.~Parisi and  M.~A.~Virasoro,  {\em
Spin Glass Theory  and Beyond} (World Scientific, Singapore 1987).

\bibitem{PB2} G.~Parisi, {\sl Field Theory, Disorder and  Simulations}, World
Scientific, (Singapore
1992).

\bi{FM} S. Franz and M. M\'ezard; {\it Off equilibrium glassy dynamics: a
simple case}, LPTENS
93/39,
{\it On mean-field glassy dynamics out of equilibrium}, cond-mat 9403004,
LPT\-ENS 94/05.

 \bi{CUKU2} L. F. Cugliandolo and J.Kurchan; Phys. Rev. Lett. {\bf 71}, 93
(1993) and references
therein.

\bibitem{MPR}E. Marinari, G. Parisi and F. Ritort,  {\sl Replica Field Theory
for Deterministic Models (II): A non-random Spin Glass with glassy Behaviour}
Hep-th/9406074,   summited to J. Phys.  {\bf A} (Math. Gen.).

\bi{CS} A.C.C. Cooley and D. Sherrington, Phys. Rev. E {\bf 49} 1921 (1994)

\bi{CF} A.C.C. Cooley and S. Franz, cond-mat preprint 9406082.

\bi{PPV} G. Parisi, N. Parga and M. A. Virasoro, J. Phys. Lettres {\bf 45}
(1984) L1063.

\bi {PP} N. Parga, G. Parisi, ICTP preprint (1985), unpublished

\bi{PARGA} N. Parga J. Phys A {\bf 19} 51 (1986).

\bi {MP} S.~Cabasino, E. Marinari, G. Parisi, P. Paolucci, J. Phys. {\bf A}
(Math. Gen.) {\bf 21}
(1988) 4201.

\bi{PV} G. Parisi and  M. A. Virasoro,  J.~Phys. {\bf 50} (1989) 3317-3329.

\bi{P1} G. Parisi  J.~Physique {\bf 51} (1990) 1595-1606.

 \bi{TAP} J. Touless, P.W. Anderson and R. G. Palmer Phil. Magaz. {\bf 35},
137 (1977).

\bibitem{CKP} L. F. Cugliandolo, J. Kurchan and G. Parisi,{\sl Off Equilibrium
Dynamics and Aging in  Unfrustrated Systems}, cond-mat preprint (1994) and
references therein.

\bi{CU} L. Cugliandolo Tor Vergata thesis (unpublished) (1994)

\bi{BBP} R. Biferale, R. Benzi, G. Parisi, Europhys. Lett. {\bf 6}, 6 (1992).

\bi{MIG} S. Migdal, Phys. Rev {\bf E 5}, 123 (1992).
 \end{thebibliography}
\end{document}